\documentclass[aps,prb,reprint,amsmath,amssymb,superscriptaddress]{revtex4-2}

\usepackage{graphicx}
\usepackage{dcolumn}
\usepackage{bm}
\usepackage{newunicodechar}
\usepackage{upgreek}
\usepackage[version=4]{mhchem} 
\usepackage{xcolor}
\usepackage{soul}
\renewcommand{\selectlanguage}[1]{}

\newunicodechar{α}{\ensuremath{\upalpha}}
\newunicodechar{β}{\ensuremath{\beta}}
\newunicodechar{λ}{\ensuremath{\lambda}}
\newunicodechar{μ}{\ensuremath{\upmu}}
\newunicodechar{θ}{\ensuremath{\theta}}
\newunicodechar{φ}{\ensuremath{\phi}}
\newunicodechar{χ}{\ensuremath{\chi}}
\newunicodechar{ε}{\ensuremath{\varepsilon}}
\newunicodechar{−}{\ensuremath{-}}
\newunicodechar{≤}{\ensuremath{\leq}}
\newunicodechar{≥}{\ensuremath{\geq}}
\newunicodechar{⋅}{\ensuremath{\cdot}}

\begin{document}

\preprint{APS/123-QED}

\title{Three-dimensional polarization of thermal fields at arbitrary distances from thin-film emitters}

\author{Xiaohong Zhang}
\email{xzhang926@gatech.edu}
\affiliation{School of Electrical and Computer Engineering, Georgia Institute of Technology, Atlanta, Georgia 30332, USA}

\author{Wenshan Cai}
\affiliation{School of Electrical and Computer Engineering, Georgia Institute of Technology, Atlanta, Georgia 30332, USA}

\author{Zhuomin M. Zhang}
\email{zhuomin.zhang@me.gatech.edu}
\affiliation{George W. Woodruff School of Mechanical Engineering, Georgia Institute of Technology, Atlanta, Georgia 30332, USA}

\date{\today}

\begin{abstract}
Recent advances have shown that thermal radiation can exhibit strong polarization in both the near and far fields, depending on the structure, material properties, and supported electromagnetic modes. Most existing studies of thermal-radiation polarization, however, focus on paraxial emission in the far field, where the polarization state can be described using conventional two-dimensional (2D) Stokes parameters. Polarization of non-paraxial thermal radiation, which is of significant importance for photonic and energy applications, remains insufficiently studied. For non-paraxial radiation, the electromagnetic field has three nonzero vector components, requiring a fully three-dimensional (3D) description of polarization. This work develops a generalized theoretical method for analyzing the polarization of thermal radiation at arbitrary distances from an emitter, under non-paraxial conditions. Fluctuation-dissipation theory is employed to determine the $3\times3$ coherence matrix of the thermally emitted electromagnetic field, and generalized 3D Stokes parameters are then obtained from the coherence matrix. Such a method is applied to planar slabs composed of isotropic, anisotropic, and nonreciprocal materials. The results reveal distinct polarization behavior across the near- and far-field regimes. In the near field, polarization is strongly influenced by evanescent waves and polariton dispersion, whereas far-field polarization is governed primarily by polarization-dependent radiative properties, including differences between transverse-electric (TE) and transverse-magnetic (TM) emission. The slab thickness further modifies these polarization characteristics by controlling the supported modes and interference effects. In addition, circular polarization, which vanishes in reciprocal materials, is observed in n-type doped InAs, whose non-reciprocity is by an external magnetic field. This work provides a unified approach for characterizing the polarization of thermal electromagnetic fields from the extreme near field to the far field.
\end{abstract}

\maketitle

\section{\label{sec:1}INTRODUCTION}

For a long time, thermal radiation has been considered to be broadband and weakly polarized because of the stochastic nature of thermally fluctuating currents within an emitter. Recently, however, numerous studies have shown the possibility of producing narrow-band and/or polarized thermal radiation through several approaches, including thermal metasurfaces~\cite{overvig_thermal_2021,wang_observation_2023,sun_full_2025,zhang_dual-polarized_2026}, anisotropic or polar materials~\cite{wu_spin-selective_2024,wu_narrowband_2025,enders_mid-infrared_2025}, and nanocarbon materials ~\cite{zacheo_efficient_2024,bharadwaj_thermal_2024,nakagawa_microemitter-based_2022}.  The generation of polarized thermal radiation through the intrinsic optical properties of the emitter, without requiring external polarization optics or complex nanofabrication, is of particular interest. 

Although extensive theoretical analyses describing the polarized thermal emission have been reported, most of them characterize the far-field emission indirectly through the polarized emissivities using Kirchhoff's law of thermal radiation. Such a treatment is particularly convenient when the radiation can be represented as a paraxial beam with a well-defined 2D transverse polarization state. However, for non-paraxial scenarios, particularly in the near field of the emitter, the electromagnetic field can have nonzero components in all three spatial directions, and the conventional 2D description becomes insufficient in this regime. Although non-paraxial far-field emission can still be analyzed by integrating the contributions from different propagation directions~\cite{zhang_nanomicroscale_2020}, such an approach still does not directly reveal the local 3D polarization state. Moreover, a description based solely on far-field emissivity cannot capture the contribution from evanescent waves that govern thermal radiation close to the emitter. A representative example is radiation above a semi-infinite planar slab: While its far-field thermal radiation is a plane wave and can be analyzed using a conventional 2D method, its near-field polarization can be strongly modified by surface waves and waveguide modes. Therefore, a generalized method for polarimetric analysis that can be applied continuously from the near field to the far field is desirable.

In this work, such a method is developed and applied to thermal emission from planar material slabs. Fluctuation-dissipation theorem (FDT) is used to determine the electromagnetic correlations of the thermal emission in the form of a $3\times3$ optical coherence matrix. The polarization state is subsequently characterized using the 3D extension of 2D Stokes parameters model, proposed by Setälä et al.~\cite{setala_degree_2002-1,setala_degree_2002}, by projecting the coherence matrix onto the Gell-Mann matrices. This work extends the previous analysis in Ref.~\cite{setala_degree_2002} on thermal radiations to a planar slab with finite thickness and focuses on specific polarization states, using a more generalized FDT method which applies to different geometries, in principle. Several types of materials are considered for their unique optical properties, including isotropic materials (ZnSe and \ce{SiO2}; Sec.~\ref{sec:isotropic}), anisotropic materials (\(\alpha\)-\ce{MoO3} and \(\beta\)-\ce{Ga2O3}; Sec.~\ref{sec:anisotropic}), and a nonreciprocal magneto-optical (InAs; Sec.~\ref{sec:nonreciprocal}). For isotropic materials, an imbalance between the in/out-of-plane polarization is observed. Surface phonon polaritons (SPhPs) are found to contribute mainly to the polarization in the near field, while the difference between TE and TM emissivities contributes to the polarization in the non-paraxial far field. For anisotropic materials, additional linear-polarization components emerge in the near field and are strongly dependent on the distance and wavelength, which is attributed to hyperbolic phonon polariton (HPhP) modes. The comparison between isotropic and anisotropic materials, with or without polariton modes, shows that SPhP and HPhP play different roles in the near-field polarization due to their different dispersion relations, while the slab thickness has more impact on the polarization in the non-paraxial far field. Besides reciprocal materials, non-reciprocal material is also investigated. For example, in n-type doped InAs, a circular-polarization component, which vanishes otherwise in a reciprocal material, appears when reciprocity is broken by an external magnetic field. These examples collectively demonstrate that the method presented in this work provides a unified description of thermal-radiation polarization across different materials and observation distances.

\section{\label{sec:2} METHODS}
\subsection{\label{sec:FDT}	Fluctuational electrodynamics}

\begin{figure}
    \centering
    \includegraphics[width=0.7\linewidth]{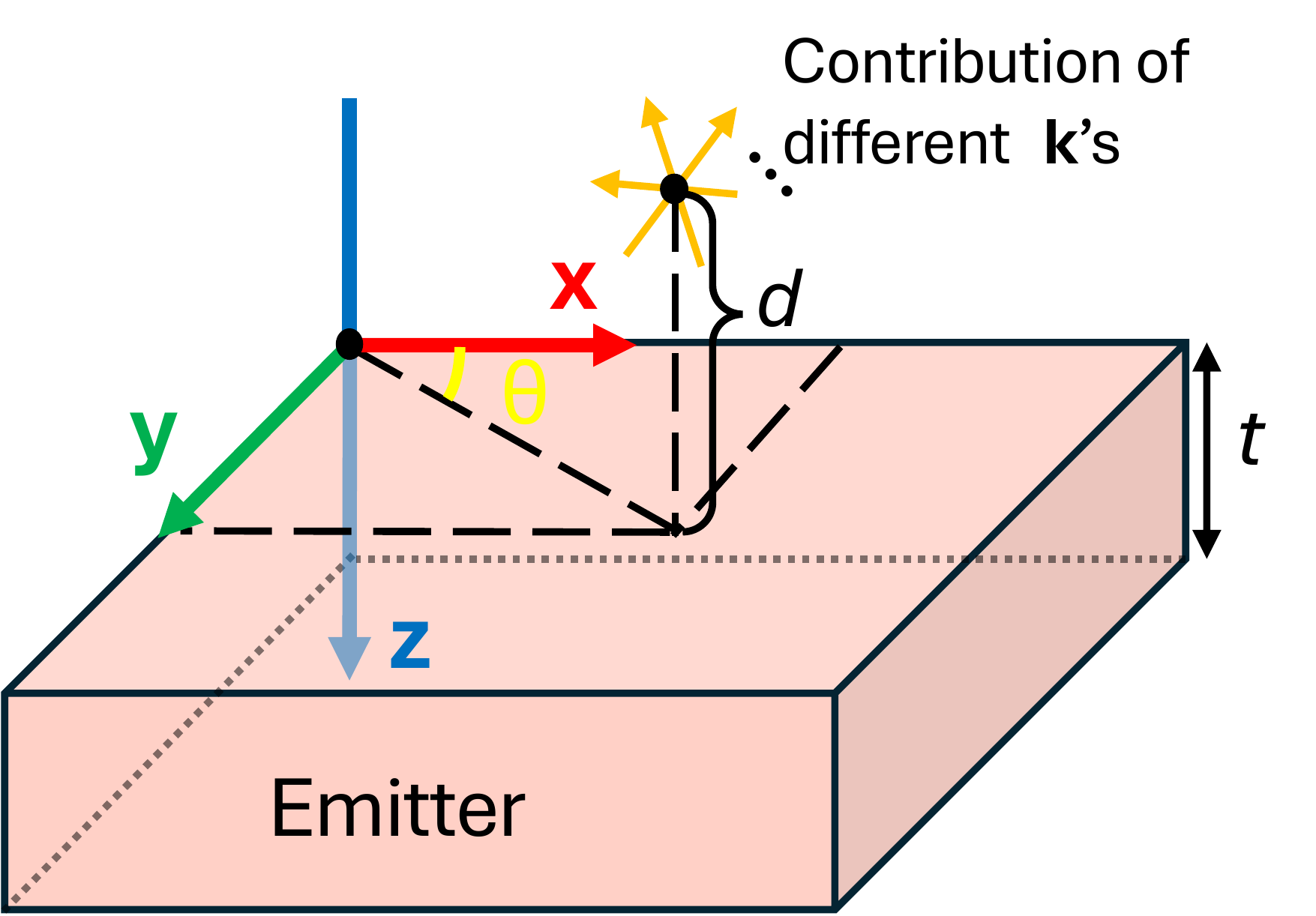}
    \caption{Coordinate setup of the calculation and the illustration of the contribution of different wavevector \textbf{k}, under non-paraxial conditions.}
    \label{fig:setup}
\end{figure}

Thermally induced microscopic current fluctuations within a material can be modeled using FDT. In the 1950s, Rytov~\cite{rytov_theory_1959} formulated the original theory and established the foundation of fluctuational electrodynamics, and later in the 1970s, Polder and van Hove extended the theory to NFRHT~\cite{polder_theory_1971}. FDT provides a general description of the statistical properties of thermal electromagnetic fields at arbitrary distances from an emitter, using a 3×3 coherence matrix that contains the correlations among the vector components of the electromagnetic field. The polarization state can therefore be determined from the coherence matrix using 3D generalized Stokes parameters, introduced in Sec.~\ref{sec:3D Stokes}. Direct application of FDT requires knowledge of the complete scattering matrix of the system, which can lead to high calculation complexity. For planar structures (referring to Fig.~\ref{fig:setup}) , an indirect method developed by Bimonte and Santamato ~\cite{bimonte_general_2007} has made the calculation simpler by assuming that the emitter and the surrounding environment are at the same temperature, i.e. global thermal equilibrium, and to calculate the case of a free-standing slab, a tiny loss can be introduced to the air behind the emitter. Using this method, the radiated field into the free space can be isolated from the total radiated field. In this section, a brief derivation of the coherence matrix is provided, and detailed derivations and numerical implementations can be found in Refs.~\cite{zhang_optical_2026,yang_polarimetric_2022,khandekar_thermal_2019}.

According to source-field relationship, the electric field \textbf{E} at an observation point \textbf{r} can be written as the spatial response of the fluctuating current density source \textbf{J}
\begin{equation}
    \label{EGJ}
    \mathbf{E}(\mathbf{r},\omega)=\mathrm{i}\omega\mu_0\int_V\bar{\bar{G}}(\mathbf{r},\mathbf{r'},\omega)\mathbf{J}(\mathbf{r'},\omega)d^3\mathbf{r'}, 
\end{equation}
where i is the unit imaginary number, $\omega$ is the angular frequency, and $\mu_0$ is the vacuum permeability. $\bar{\bar{G}}$ is the dyadic Green’s function of a planar slab. $\mathbf{r'}$ is the source position vector and ranges over the entire source region. In principle, Eq.~\eqref{EGJ} is not limited to planar geometries and also applies to other geometries, such as a sphere~\cite{yeh_dyadic_1963}, as long as the Green's function is known.

For thermal emission, the ensemble-average electric field can be described by coherence matrix $\bar{\bar{W}}$, which is evaluated by subtracting the far-field blackbody radiation component, $\mathrm{\mathbf{E}_{vac}}$, from the total electric field surrounding a thermal emitter in global thermal equilibrium $\mathrm{\mathbf{E}_{ge}}$~\cite{bimonte_general_2007}
\begin{equation}\label{matrix_substract}
    \begin{aligned}
    \bar{\bar{W}}&=\left\langle\mathbf{E}(\mathbf{r},\omega)\otimes\mathbf{E}^{*}(\mathbf{r},\omega)\right\rangle\\
    &=\left\langle\mathbf{E}_\mathrm{ge}(\mathbf{r},\omega)\otimes\mathbf{E}_\mathrm{ge}^{*}(\mathbf{r},\omega)\right\rangle
    \!-\!\left\langle\mathbf{E}_\mathrm{vac}(\mathbf{r},\omega)\otimes\mathbf{E}_\mathrm{vac}^{*}(\mathbf{r},\omega)\right\rangle,
    \end{aligned}
\end{equation}
and the closed form of the coherence matrix is
\begin{equation}
    \label{matrix}
    \begin{aligned}
    \bar{\bar{W}} = \begin{bmatrix}
    \left\langle E_{x}E_{x}^{\dagger} \right\rangle & \left\langle E_{x}E_{y}^{\dagger} \right\rangle & \left\langle E_{x}E_{z}^{\dagger} \right\rangle \\
    \left\langle E_{y}E_{x}^{\dagger} \right\rangle & \left\langle E_{y}E_{y}^{\dagger} \right\rangle & \left\langle E_{y}E_{z}^{\dagger} \right\rangle \\
    \left\langle E_{z}E_{x}^{\dagger} \right\rangle & \left\langle E_{z}E_{y}^{\dagger} \right\rangle & \left\langle E_{z}E_{z}^{\dagger} \right\rangle
    \end{bmatrix},
    \end{aligned}
\end{equation}
where each element of the matrix has units of $\mathrm{V^2\cdot s\cdot m^{-1}\cdot rad^{-1}}$ as they are spectral field correlations. The autocorrelation of $\mathrm{\mathbf{E}_{ge}}$ is given by
\begin{equation}
    \label{Ege}
    \begin{aligned}
    \langle \mathbf{E_{\mathrm{ge}}}(\mathbf{r},\omega)\otimes\mathbf{E^*_{\mathrm{ge}}}(\mathbf{r},\omega) \rangle
    = -\omega^2\mu_0^2\iint_V\bar{\bar{G}}(\mathbf{r},\mathbf{r'},\omega)
    \\ \times \langle \mathbf{J}(\mathbf{r'},\omega) 
    \otimes \mathbf{J}^*(\mathbf{r''},\omega) \rangle\bar{\bar{G}}^\dagger(\mathbf{r},\mathbf{r''},\omega)d^3\mathbf{r'}d^3\mathbf{r''}.
    \end{aligned}
\end{equation}
Here, $\bar{\bar{A}} \otimes \bar{\bar{B}} = \bar{\bar{A}}\bar{\bar{B}}^\mathrm{T}$ represents the tensor product or outer product, with the superscript T as matrix transpose. \(\dagger\) denotes the Hermitian conjugate. The first kind of FDT shows that the correlation between two current density sources is nonzero only if they are at the same location~\cite{zhang_nanomicroscale_2020,rytov_theory_1959}
\begin{equation}
    \label{JJ}
    \begin{aligned}
    \left\langle \mathbf{J}\left( \mathbf{r}',\omega \right) \otimes \mathbf{J}^{*}\left( \mathbf{r}'',\omega \right) \right\rangle &= \frac{4}{\pi}\omega\varepsilon_{0}\Theta(\omega,T) \\
    &\times \frac{\bar{\bar{\varepsilon}}(\omega) - {\bar{\bar{\varepsilon}}}^{\dagger}(\omega)}{2\mathrm{i}}\delta\left( \mathbf{r}' - \mathbf{r}'' \right),
    \end{aligned}
\end{equation}
where
\(\Theta(\omega,T) = \frac{\hbar\omega}{2} + \hbar\omega/(e^{\hbar\omega/k_\mathrm{B}T} - 1)\)
is the mean energy of Planck's oscillator, with \(k_\mathrm{B}\) the Boltzmann
constant, \(\hbar\) the reduced Planck constant, and \emph{T} the
temperature. All examples in this work have a uniform temperature. \(\varepsilon_{0}\) and
\(\bar{\bar{\varepsilon}}\) are the vacuum permittivity 
and the relative permittivity tensor of the medium. \(\delta\) denotes the Dirac delta. Hence, Eq.~\eqref{Ege} can be simplified accordingly
\begin{equation}
    \label{simp_Ege}
    \begin{aligned}
    \langle \mathbf{E_{\mathrm{ge}}}(\mathbf{r},\omega)\otimes\mathbf{E^*_{\mathrm{ge}}}(\mathbf{r},\omega) \rangle
    =-\frac{4\omega\mu_0\Theta(\omega,T)}{\pi}
    \\\times\frac{\omega^2}{c^2}\int_V\bar{\bar{G}}(\mathbf{r},\mathbf{r'},\omega)\frac{\bar{\bar{\varepsilon}}(\omega) - \bar{\bar{\varepsilon}}^\dagger(\omega)}{2\mathrm{i}}\bar{\bar{G}}^{\dagger}(\mathbf{r},\mathbf{r'},\omega)d^3\mathbf{r'},
    \end{aligned}
\end{equation}
where $c=(\mu_0\varepsilon_0)^{-1/2}$ is the speed of light in vacuum. An identity of the Green's function can be applied to further simplify the integral part of Eq.~\eqref{simp_Ege}~\cite{dung_three-dimensional_1998,zhang_optical_2026},
so that it becomes
\begin{equation}
    \label{simp2_Ege}
    \begin{aligned}
    \langle \mathbf{E_{\mathrm{ge}}}(\mathbf{r},\omega)\otimes\mathbf{E^*_{\mathrm{ge}}}(\mathbf{r},\omega) \rangle = 2\mathrm{i}\omega\mu_0\Theta(\omega,T)
    \\\times\frac{\bar{\bar{G}}\left( \mathbf{r},\mathbf{r},\omega \right) - {\bar{\bar{G}}}^{\dagger}\!\left( \mathbf{r},\mathbf{r},\omega \right)}{\pi}.
    \end{aligned}
\end{equation}
The vacuum contribution can be modeled using a plane-wave expansion, since the boundary blackbody is assumed to be infinitely far away from the emitter. Using the coordinate system as in Fig.~\ref{fig:setup}, the \textit{k}-space components of the vacuum electric field are~\cite{zhang_optical_2026}
\begin{equation}
    \label{Evac}
    \begin{aligned}
    \mathbf{E}_\mathrm{vac}^{s}(z,\mathbf{k}_{\parallel},\omega)&=a_s\left[e^{\mathrm{i}k_{0z}z}\widehat{\mathbf{s}}_{+}+e^{-\mathrm{i}k_{0z}z}\left(r_{ss}\widehat{\mathbf{s}}_{-}+r_{sp}\widehat{\mathbf{p}}_{-}\right)\right],\\
    \mathbf{E}_\mathrm{vac}^{p}(z,\mathbf{k}_{\parallel},\omega)&=a_p\left[e^{\mathrm{i}k_{0z}z}\widehat{\mathbf{p}}_{+}+e^{-\mathrm{i}k_{0z}z}\left(r_{ps}\widehat{\mathbf{s}}_{-}+r_{pp}\widehat{\mathbf{p}}_{-}\right)\right],
\end{aligned}
\end{equation}
where $k_{\parallel}$ is the transverse wavevector in the \textit{xy} plane, $k_{0z}=(k_0^2-k_{\parallel}^2)^{1/2}$ is the longitudinal wavevector along \textit{z}, and $r_{ij}$ is the reflection coefficient from incident \textit{i} polarization (\textit{s} or \textit{p}) to reflected \textit{j} polarization (\textit{s} or \textit{p}), with the \textit{s} polarization defined along the \textit{y} direction. The coefficient $a_{s(p)}$ is the amplitude of an $s$- or $p$-polarized wave. Its autocorrelation is defined through the amplitude of the field emitted from a blackbody:
\begin{equation}
    \label{aa}
    \left\langle a_i a_j^*\right\rangle=\delta_{ij}\delta\!\left(\mathbf{k}_{\parallel}-\mathbf{k}_{\parallel}'\right)4\pi\frac{\mu_0\omega}{k_{0z}}\Theta(\omega,T),
    \end{equation}
where $\delta_{ij}$ is the Kronecker delta. The vacuum electric field in real space can then be obtained by performing a Fourier transform. For simplicity, the detailed derivation of the Green's function and evaluation techniques in the \textit{k}-space are covered in Ref.~\cite{zhang_optical_2026,yang_polarimetric_2022}.

\subsection{\label{sec:3D Stokes}	Three-dimensional Stokes Parameters}
The polarization state of an electromagnetic field with a planar wavefront can be mapped onto four Pauli matrices and completely described by a set of four parameters, including for partially polarized fields~\cite{born_principles_1999}. These parameters enable the polarimetric analysis of far-field thermal radiation, offering the foundation for polarization control as well. For general non-paraxial conditions, including the near field, the electromagnetic field no longer has a well-defined planar wavefront, and therefore its polarization cannot be described by conventional 2D Stokes parameters. A generalized 3D description can instead be constructed by expanding the $3\times3$ coherence matrix in a basis formed by the identity matrix and the eight Gell-Mann matrices~\cite{gell-mann_symmetries_1962,setala_degree_2002-1}
\begin{equation}
    \label{Gell-Mann}
    \begin{aligned}
    &\bar{\bar{\lambda}}_0 = 
    \begin{bmatrix}
    1 & 0 & 0 \\
    0 & 1 & 0 \\
    0 & 0 & 1
    \end{bmatrix},
    \bar{\bar{\lambda}}_1 = 
    \begin{bmatrix}
    0 & 1 & 0 \\
    1 & 0 & 0 \\
    0 & 0 & 0
    \end{bmatrix},
    \bar{\bar{\lambda}}_2 = 
    \begin{bmatrix}
    0 & -\mathrm{i} & 0 \\
    \mathrm{i} & 0 & 0 \\
    0 & 0 & 0
    \end{bmatrix},\\
    &\bar{\bar{\lambda}}_3 = 
    \begin{bmatrix}
    1 & 0 & 0 \\
    0 & -1 & 0 \\
    0 & 0 & 0
    \end{bmatrix},
    \bar{\bar{\lambda}}_4 = 
    \begin{bmatrix}
    0 & 0 & 1 \\
    0 & 0 & 0 \\
    1 & 0 & 0
    \end{bmatrix},
    \bar{\bar{\lambda}}_5 = 
    \begin{bmatrix}
    0 & 0 & -\mathrm{i} \\
    0 & 0 & 0 \\
    \mathrm{i} & 0 & 0
    \end{bmatrix},\\
    &\bar{\bar{\lambda}}_6 = 
    \begin{bmatrix}
    0 & 0 & 0 \\
    0 & 0 & 1 \\
    0 & 1 & 0
    \end{bmatrix},
    \bar{\bar{\lambda}}_7 = 
    \begin{bmatrix}
    0 & 0 & 0 \\
    0 & 0 & -\mathrm{i} \\
    0 & \mathrm{i} & 0
    \end{bmatrix},
    \bar{\bar{\lambda}}_8 = \frac{1}{\sqrt{3}}
    \begin{bmatrix}
    1 & 0 & 0 \\
    0 & 1 & 0 \\
    0 & 0 & -2
    \end{bmatrix},
    \end{aligned}
\end{equation}
and the coherence matrix can be decomposed accordingly by
\begin{equation}
\label{decompose}
    \bar{\bar{W}}=\frac{1}{3}\sum^8_{n=0}\Lambda_n\bar{\bar{\lambda}}_n^*,
\end{equation}
and therefore the 3D Stokes parameters are
\begin{equation}
    \label{3D_Stokes}
    \begin{array}{ll}
    \Lambda_0 = \bar{\bar{W}}_{xx}+\bar{\bar{W}}_{yy}+\bar{\bar{W}}_{zz},
    &\Lambda_1 = \frac{3}{2}(\bar{\bar{W}}_{xy}+\bar{\bar{W}}_{yx}),\\[3pt]
    \Lambda_2 = \frac{3}{2}\mathrm{i}(\bar{\bar{W}}_{xy}-\bar{\bar{W}}_{yx}),
    &\Lambda_3 = \frac{3}{2}(\bar{\bar{W}}_{xx}-\bar{\bar{W}}_{yy}),\\[3pt]
    \Lambda_4 = \frac{3}{2}(\bar{\bar{W}}_{xz}+\bar{\bar{W}}_{zx}),
    &\Lambda_5 = \frac{3}{2}\mathrm{i}(\bar{\bar{W}}_{xz}-\bar{\bar{W}}_{zx}),\\[3pt]
    \Lambda_6 = \frac{3}{2}(\bar{\bar{W}}_{yz}+\bar{\bar{W}}_{zy}),
    &\Lambda_7 = \frac{3}{2}\mathrm{i}(\bar{\bar{W}}_{yz}-\bar{\bar{W}}_{zy}),\\[3pt]
    \Lambda_8 = \frac{\sqrt{3}}{2}(\bar{\bar{W}}_{xx}+\bar{\bar{W}}_{yy}-2\bar{\bar{W}}_{zz}),
    \end{array}
\end{equation}
and they are purely real because the coherence matrix is Hermitian. Among these Stokes parameters, $\Lambda_0$ represents the total electric-field correlation. $\Lambda_1,\ \Lambda_4,$ and $\Lambda_6$ represent the $45^\circ/135^\circ$ linear polarization components in the $xy,\ xz$, and $yz$ planes, respectively. $\Lambda_2,\ \Lambda_5,$ and $\Lambda_7$ represent the left-/right-hand circular polarization (LCP/RCP) components in the $xy,\ xz,$ and $yz$ planes, respectively. $\Lambda_3$ represents the $0^\circ/90^\circ$ linear polarization components in the $xy$ plane, and $\Lambda_8$ describes the imbalance between the combined in-plane components and the out-of-plane component. The use of $\Lambda_8$ is to account for the fact that a 3D electromagnetic field cannot be arbitrarily decomposed into a fully polarized field and a fully unpolarized field, because the 3D unpolarized field may have different intensities in different coordinate planes~\cite{setala_degree_2002-1}. Fundamentally speaking, the eigenvalues of the unpolarized field eigenvectors can be forced the same in 2D cases but cannot in the generic 3D cases.

Analogous to the 2D case, the degree of individual polarization can be defined by taking the ratio between $\Lambda_i$ and $\Lambda_0$, named normalized Stokes parameters, where $i\neq0$, with $\Lambda_{1-7}/\Lambda_0$ ranging from -3/2 to 3/2 and $\Lambda_8/\Lambda_0$ ranging from $-\sqrt{3}$ to $\sqrt{3}$. By taking such a ratio, the temperature dependency caused by oscillator term, $\Theta(\omega,T)$, is eliminated. It should be emphasized that $\Lambda_i/\Lambda_0$ is different from the conventional 2D degree of linear/circular polarization, as the contribution from $\Lambda_8$ is isolated.

Furthermore, the total degree of polarization (DoP) can be defined as
\begin{equation}
    \label{TotalDoP3D}
    \mathrm{DoP_{3D}}=\sqrt{\frac{\displaystyle\sum_{i=1}^8\Lambda^2_i}{3\Lambda_0^2}}=\sqrt{\frac{3}{2} \left(\frac{\operatorname{tr} (\bar{\bar{W}}^2) }{\operatorname{tr}^2(\bar{\bar{W}})} - \frac{1}{3}\right)},
\end{equation}
where ``tr" stands for trace, and the factor of $1/\sqrt{3}$ is for normalization, so that $\mathrm{DoP}$ lies between 0 and 1. In the rest of this paper, $\Lambda_i/\Lambda_0$ is mainly used to analyze each nonzero polarization state in thermal radiation, while $\mathrm{DoP_{3D}}$ is used to describe the overall degree of polarization.

\section{\label{sec:3} Results and Discussion}
In this section, polarimetric analysis is performed on three different material types: (1) Isotropic, (2) Anisotropic (hyperbolic), and (3) Nonreciprocal. By comparing different materials within and across each category, unique and intriguing phenomena of the thermal radiation of those materials are observed. In isotropic materials, no specific polarization is observed at any distance, except in/out-of-plane polarization imbalance, $\Lambda_8$, due to isotropy of the planar geometry and the emissivity difference between the TE and TM waves. In hyperbolic materials, strong near-field polarization is observed, whose polarization angle can be explained by the polariton modes and iso-frequency contours (IFCs). In the nonreciprocal material, n-type doped InAs, the calculation results show circular polarization in the near field, which does not exist in reciprocal materials. In general, this method reveals the specific polarization features in each material category.

\subsection{\label{sec:isotropic} Isotropic materials}
Since very little research has been done on the polarization of the thermal near field and under non-paraxial conditions, it is natural to start with the simplest case: an isotropic planar slab. The radiative behavior of an isotropic planar slab can also be used as a reference when compared with other types of materials. Two commonly used materials are considered for mid-infrared (mid-IR) optical or thermal applications, ZnSe and \ce{SiO2} (fused silica). Those two materials are chosen for their different optical properties in the mid-IR region. ZnSe has a very stable refractive index ($n\approx2.3\text{-}2.4$) and extremely low loss ($\kappa\leq1\times10^{-6}$) across the mid-IR spectrum~\cite{querry_optical_1987}, while \ce{SiO_2} is highly absorptive around 9 μm and 21 μm due to phonon absorptions~\cite{philipp_silicon_1985}. As will be shown later, the  excitation of SPhP in \ce{SiO2} can affect the degree of polarization greatly.

\begin{figure}[t]
    \centering
    \includegraphics[width=1\linewidth]{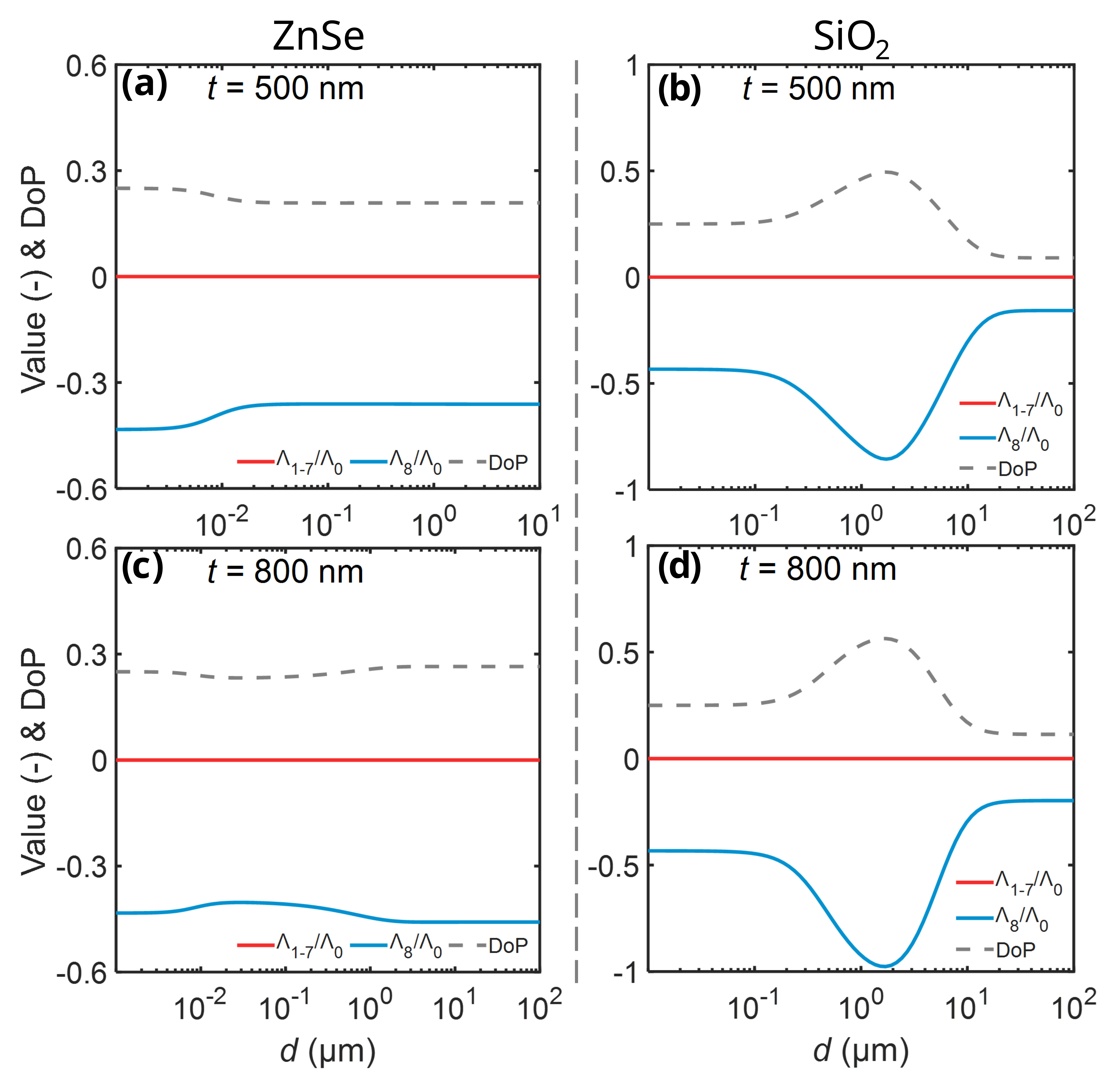}
    \caption{The normalized Stokes parameters, $\Lambda_{1\text{-}7}/\Lambda_0$ and $\Lambda_8/\Lambda_0$, and the total DoP of the single slab ZnSe [(a) and (c)] and \ce{SiO2} [(b) and (d)] with different thicknesses, at $\lambda = 9$  μm.}
    \label{fig:ZnSe Stokes}
\end{figure}

To start with, a planar slab of ZnSe/\ce{SiO2} with a thickness of \textit{t} is configured as shown in Fig.~\ref{fig:setup}. The air-material interface is at $z=0$. The observation point is located a distance \textit{d} from the emitter, corresponding to $z=−d$. The 3D Stokes parameters are calculated at wavelength λ = 9 μm and normalized to $\Lambda_0$, as shown in Fig.~\ref{fig:ZnSe Stokes}. All Stokes parameters, except $\Lambda_0$ and $\Lambda_8$, are zero due to the preservation of rotational symmetry of the isotropic slab. The nonzero $\Lambda_8$ is caused by the difference between the TE and TM emissivities, which is due to different reflectance and transmittance, according to Kirchhoff’s law. Because $\Lambda_8/\Lambda_0$ is the only nonzero Stokes component, it contributes solely to the total DoP according to Eq.(\ref{TotalDoP3D}).

\begin{figure}[!t]
    \centering
    \includegraphics[width=1\linewidth]{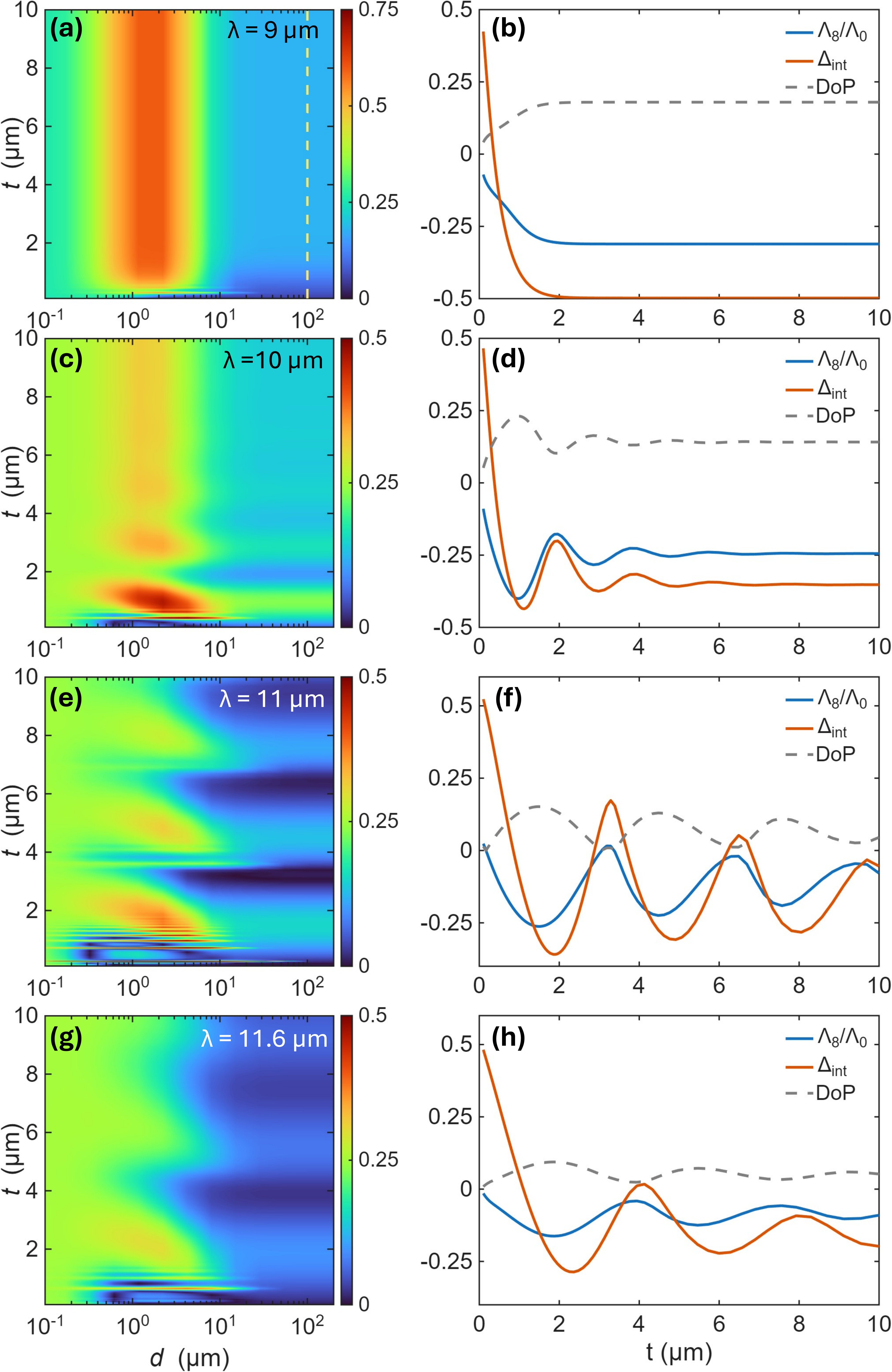}
    \caption{Left column; (a), (c), (e), and (g): Total DoP versus the \ce{SiO2} slab thickness $t$ and distance to the top surface $d$, at different wavelengths. Note that (a) uses a different range for colorbar as the magnitude of DoP is much higher than the others. Right column; (b), (d), (f), and (h): Total DoP, $\Lambda_8/\Lambda_0$, and $\Delta_{int}$ versus the slab thickness, at a fixed distance $d = 100$ μm, whose position on the plot is marked as the yellow dotted line in (a) for instance.}
    \label{fig:Delta_TETM}
\end{figure}

The DoP trends of ZnSe and \ce{SiO2} share similarities but also exhibit many differences. The DoPs of the two materials converge to the same value of 0.25 as $d\rightarrow0$, which is consistent with the universal extreme-near-field limit predicted in Ref.~\cite{setala_degree_2002}. In this extreme near field regime, the surface waves with large $k_\parallel$ values dominate the field correlations, causing the polarization behavior to become insensitive to the other modes in the material slab~\cite{lau_spatial_2007,setala_degree_2002,henkel_spatial_2000}. At larger distances, the effect of the high-$k$ surface waves starts decreasing rapidly, while other material-specific surface modes become important, such as SPhP or waveguide modes. One major difference between the polarization of emission from ZnSe and \ce{SiO2} occurs at this distance due to the SPhP mode: The DoP of the \ce{SiO2} slab exhibits local minima (magnitude maxima), while the DoP of the ZnSe slab remains nearly flat and smooth. At λ = 9 μm, ZnSe does not support any SPhP, whereas \ce{SiO2} supports SPhP. Because this mode possesses a strong $E_z$ component, it produces a large imbalance between the in/out-of-plane field correlations  and therefore increases the magnitude of $\Lambda_8/\Lambda_0$ to nearly $-1$ and the total DoP to 0.56 for $t=800$ nm, as shown in Fig.~\ref{fig:ZnSe Stokes} (d). The SPhP in \ce{SiO2} propagates equally in every direction in the $xy$ plane. Thus, there is no in-plane polarization induced.

\begin{figure}[b]
    \centering
    \includegraphics[width=1\linewidth]{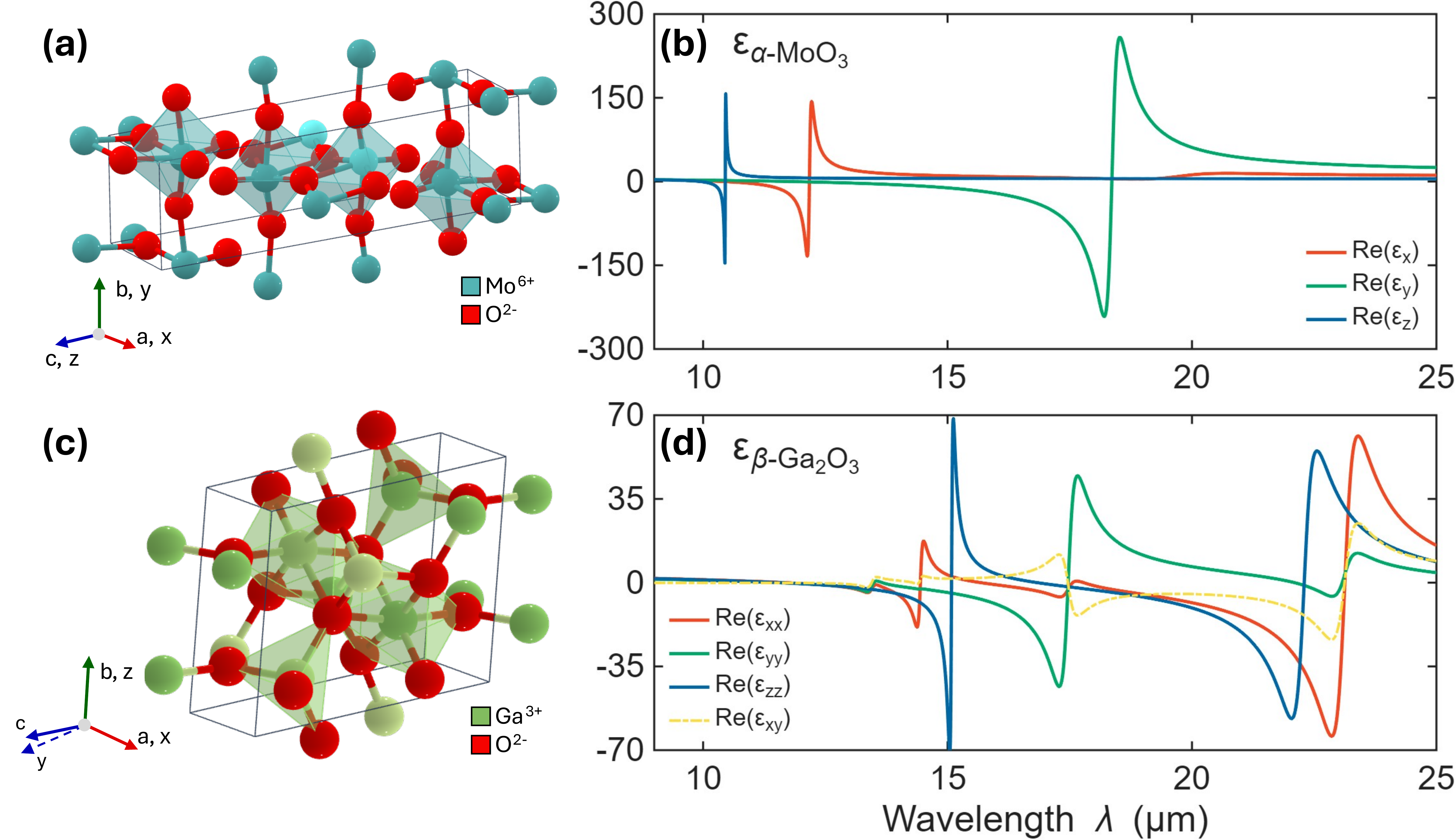}
    \caption{(a) The primitive cell of α-\ce{MoO3}. The primitive cell vectors are aligned with the Cartesian coordinate axes shown in the bottom left. (b) The real part of the dielectric function of α-\ce{MoO3} from 9 to 25 μm. (c) Primitive cell of β-\ce{Ga2O3}. The primitive cell vectors a and b are aligned with $x$ and $z$ axes, respectively. Due to monoclinic structure, vector c cannot be mapped onto the $y$ axis. (d) The real part of the dielectric function of β-\ce{Ga2O3} from 9 to 25 μm. The crystal structure illustrations are obtained from Materials Project~\cite{noauthor_materials_2020-1,noauthor_materials_2020}.}
    \label{fig:crystal}
\end{figure}
\begin{figure*}
    \centering
    \includegraphics[width=0.9\linewidth]{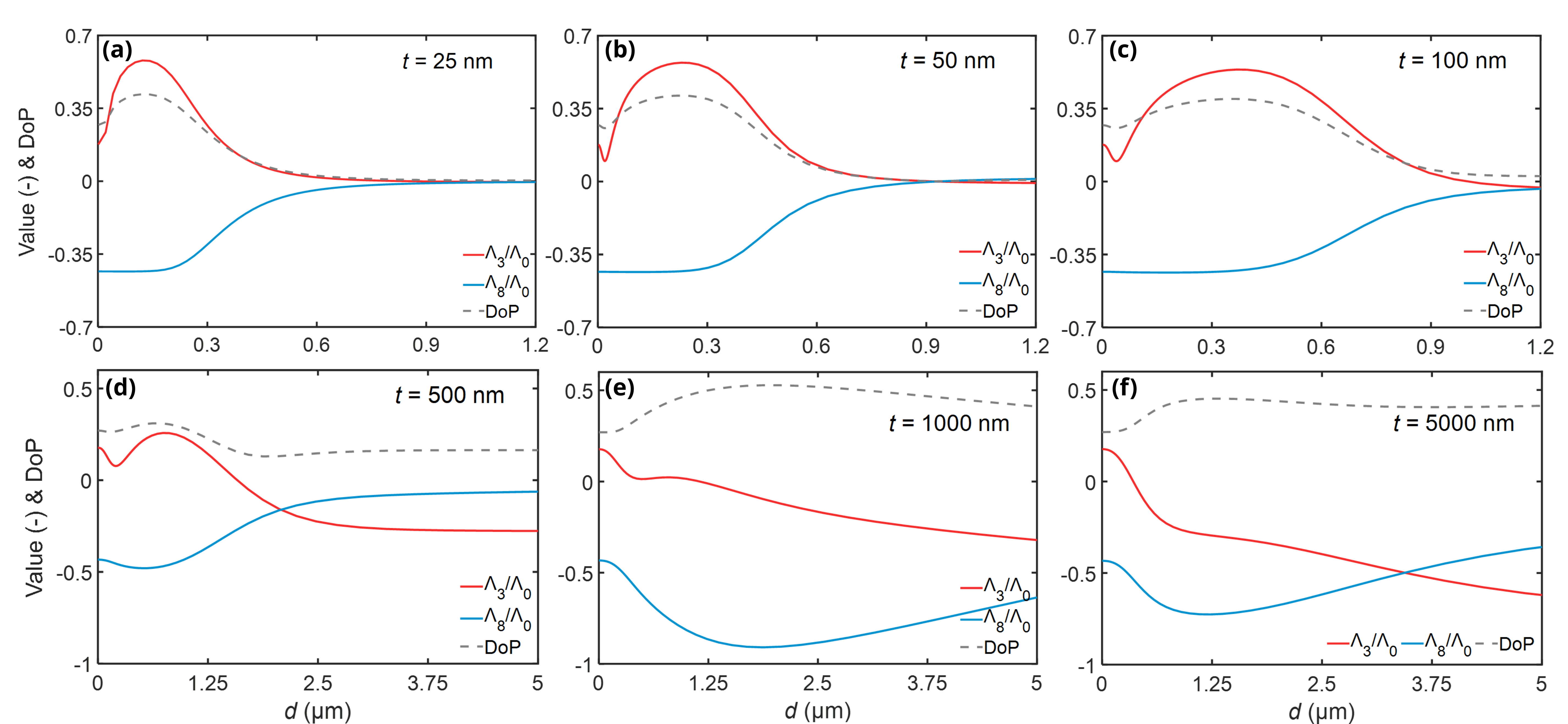}
    \caption{Normalized Stokes parameters, $\Lambda_3/\Lambda_0$ and $\Lambda_8/\Lambda_0$ (out-of-plane \textit{z}-polarization), and total DoP of the single slab α-\ce{MoO3} with different thickness at λ = 11 μm. Other Stokes parameters are zero, and are not shown. Note that for thin-thickness cases (a)-(c) the \textit{x}-axis ranges from 0 to 1.2 μm, while for large-thickness cases (d)-(f) the \textit{x}-axis ranges from 0 to 5 μm.}
    \label{fig:MoO3_Stokes}
\end{figure*}

Another difference occurs in the far field: When $d>\lambda$, the far-field limiting DoP values for both cases are different. In addition, changing the thickness of the slab also varies the far-field values, as shown in Fig.~\ref{fig:ZnSe Stokes} at $d>9$ μm. In the far field, the contribution of surface waves vanishes, and therefore only the far-field polarization response of the structure, such as polarization-dependent reflectance/transmittance and multiple reflections inside the slab, contribute to the field polarization. Since the emission is non-paraxial, the directional contributions from different wavevectors need to be taken into account, and therefore the angle-integrated emissivity difference $\Delta_{int}$ is defined as follows
\begin{equation}
    \label{eqn:diff_TE_TM}
    \Delta_{int}=\int^{\pi/2}_0(\epsilon_{\mathrm{TE}}-\epsilon_{\mathrm{TM}})\ d\theta,
\end{equation}
where, according to the Kirchhoff's law of thermal radiation, $\epsilon = \alpha = 1-\rho-\tau$, with $\epsilon$ as the emissivity, $\alpha$ the absorptivity, $\rho$ the reflectivity, and $\tau$ the transmissivity. Azimuthal angle is not considered as the material is isotropic. By comparing the $\Delta_{int}$ with DOP and normalized Stokes parameters as in Fig.~\ref{fig:Delta_TETM}, it can be clearly seen that there is a clear association between the DoP, $\Lambda_8/\Lambda_0$, and $\Delta_{int}$ in the far field, except for $\lambda=9$ μm. The potential reason behind this exception is that, at the phonon resonance wavelength, the contribution due to the surface wave contribution dominates the other effects. The loss inside the material is also very high, therefore leading to the bulk behavior of the slab and making the thin film multiple reflection model no longer valid. Therefore, the correlation between $\Delta_{int}$ and other quantities is broken. For wavelengths outside the optical phonon region, the association between $\Delta_{int}$ and DoP is clearer, verifying the contribution of the polarized emissivity difference to the polarization in the far field. It should be noted that $\Delta_{int}$ is a different physical quantity calculated using different methods from DoP and $\Lambda_8/\Lambda_0$, and therefore their magnitudes can differ and need not match exactly, while the correlation between them still holds.

The far-field DoP of \ce{SiO2} slab is also wavelength dependent. The DoP versus thickness plot has a higher damping rate at 10 μm (Fig.~\ref{fig:Delta_TETM}(d)) and 11.6 μm (Fig.~\ref{fig:Delta_TETM}(h)) as the slab thickness \textit{t} increases, because \ce{SiO2} has a higher loss at these two wavelengths ($\kappa = 0.082$ and 0.048, respectively), compared with $\kappa = 0.017$ at 11 μm(Fig.~\ref{fig:Delta_TETM}(f)). The higher loss hinders multiple reflections and interference inside the slab, therefore making the DoP and the $\Delta_{int}$ approach the values for bulk material more quickly.

\begin{figure}
    \centering
    \includegraphics[width=1\linewidth]{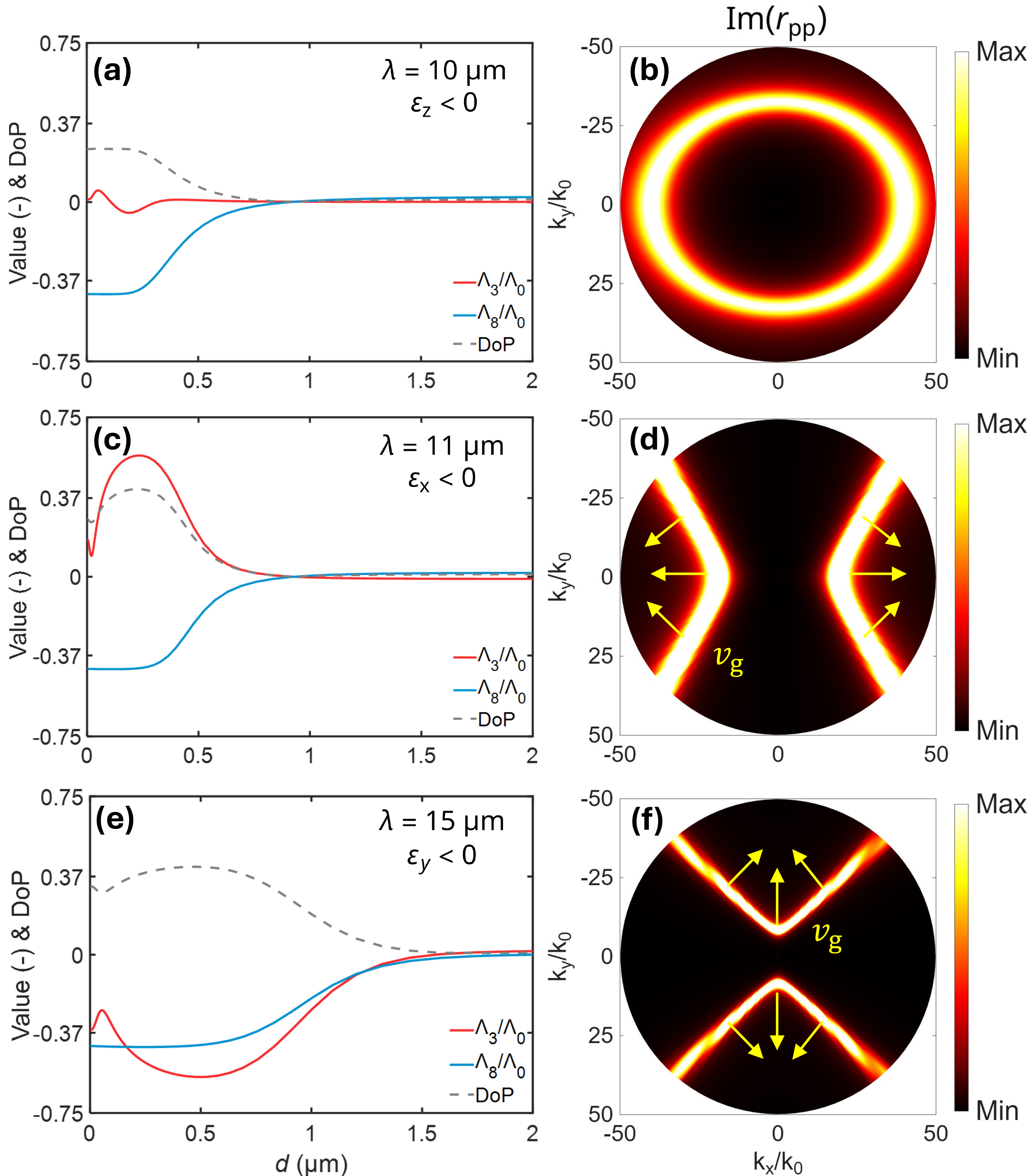}
    \caption{Normalized Stokes components of a 50-nm α-\ce{MoO3} planar slab at different wavelengths and the corresponding polariton modes, represented by their IFCs. The yellow arrows indicate the direction of the group velocity, $\mathbf{v}_g$.}
    \label{fig:MoO3_HPhP}
\end{figure}

\begin{figure*}[t]
    \centering
    \includegraphics[width=0.9\linewidth]{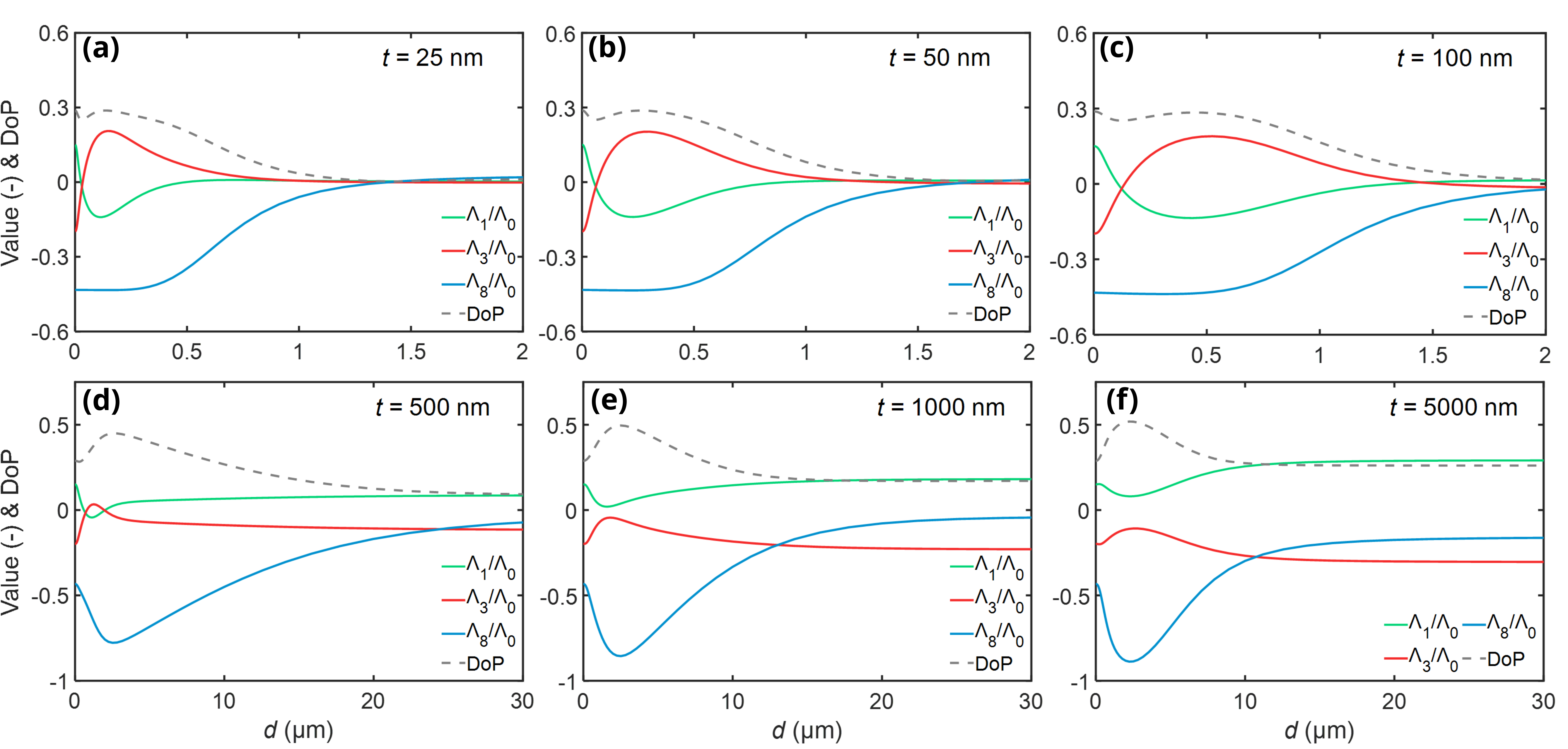}
    \caption{The Stokes parameters ($\Lambda_1$, $\Lambda_3$, and $\Lambda_8$; others are zero) and total DoP of the single slab β-\ce{Ga2O3} with different thickness at λ = 13.9 μm. Note that for thin-thickness cases (a)-(c) the $x$ axis ranges from 0 to 2 μm, while for large-thickness cases (d)-(f), the $x$ axis ranges from 0 to 30 μm.}
    \label{fig:Ga2O3_Stokes}
\end{figure*}

\subsection{\label{sec:anisotropic} Anisotropic (hyperbolic) materials}
In this section, the polarization states of thermal emission from slabs of two anisotropic material, \(\alpha\)-\ce{MoO3} and \(\beta\)-\ce{Ga2O3}, are discussed. These two materials are chosen for their intriguing hyperbolic permittivity tensors, which can support HPhPs~\cite{wang_observation_2025,zhou_engineering_2025}. Unlike an ordinary SPhP mode, HPhP is a type of polariton mode which has a constrained direction of propagation, and therefore may contribute to the near-field polarization in a more pronounced way. To be more specific, \(\alpha\)-\ce{MoO3}, whose crystal planes along the $c$ axis are bonded by van der Waals forces, has a diagonal hyperbolic permittivity tensor, while the permittivity tensor of \(\beta\)-\ce{Ga2O3} contains nonzero off-diagonal elements. The crystal structures and permittivity spectra are shown in Fig.~\ref{fig:crystal}.

To start with, an α-\ce{MoO3} slab with varying thickness is considered at λ = 11 μm, within the second Reststrahlen band as shown in Fig.~\ref{fig:MoO3_Stokes}. At this wavelength, $\operatorname{Re}(\varepsilon_x)<0$ while $\operatorname{Re}(\varepsilon_y)>0$, resulting in an in-plane hyperbolic IFC whose open branches are oriented predominantly along the $k_x$ direction. The group velocity, which is normal to the IFC, is consequently confined to a restricted range of real-space propagation directions. The hybrid longitudinal HPhP wave therefore allows the linear polarization in the near field along both the $x$ and $z$ directions, which corresponds to a positive $\Lambda_3/\Lambda_0$ and a negative $\Lambda_8/\Lambda_0$ at $d\approx0.3$ μm.

For different thicknesses of the α-\ce{MoO3} slab, the peak magnitude and the decay lengths of $\Lambda_3/\Lambda_0$ are different. At $t = 25$ nm, the maximum $\Lambda_3/\Lambda_0$ is 0.58, while at $t = 500$ nm, it reduces to 0.23. As the thickness further increases, the peak completely vanishes, leaving only the obvious features in the extreme near field and the far field. The decay length of $\Lambda_3/\Lambda_0$ increases as the thickness increases.

To further investigate the cause of linear polarization in the near field due to HPhPs, different wavelengths within different Reststrahlen bands of α-\ce{MoO3} are considered. The normalized Stokes parameters and IFCs are shown in Fig.~\ref{fig:MoO3_HPhP}. At $\lambda = 10$ μm, $\varepsilon_z<0$, no in-plane HPhPs are supported, and the IFC is an ellipse due to in-plane anisotropy. At this wavelength the in-plane permittivities are not highly anisotropic, the group velocity is thus not highly confined. The linear polarization is weak in this case, which is similar to the SPhP in \ce{SiO2} as shown in Sec.~\ref{sec:isotropic}. At $\lambda = 11$ μm, however, $\varepsilon_x<0$ makes the IFC a hyperbola in the $xy$ plane, with opening towards the $x$ axis. In this case, the group velocity is confined in the $x$ direction and therefore induces linear polarization along the $x$ and $z$ direction. Similarly, at $\lambda = 15$ μm, the hyperbolic IFC has its opening toward $y$ axis, and therefore produces polarization along the $y$ and $z$ directions, leading to negative $\Lambda_3$ and $\Lambda_8$.
\begin{figure}
    \centering
    \includegraphics[width=1\linewidth]{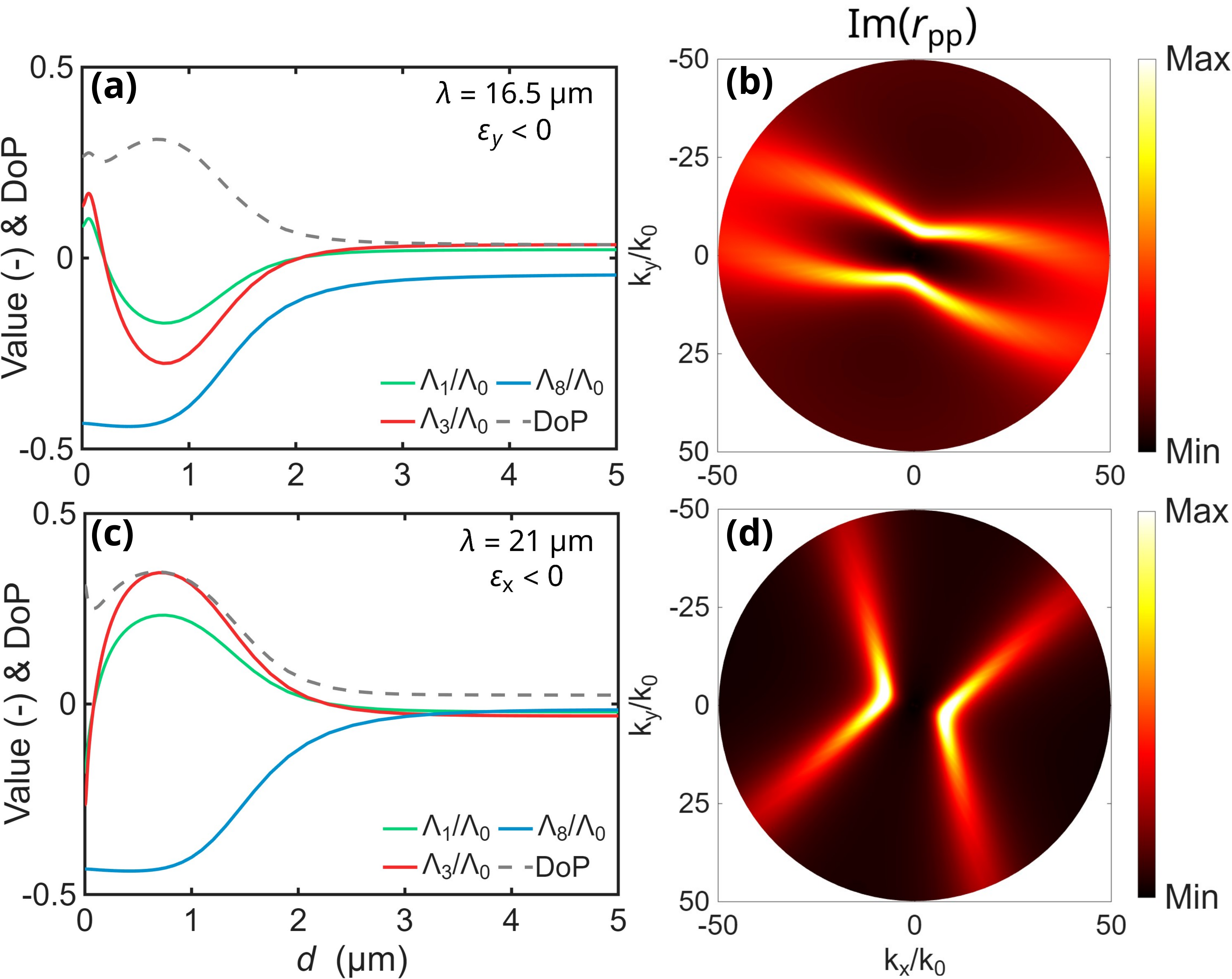}
    \caption{Normalized Stokes components of a 50-nm β-\ce{Ga2O3} planar slab at different wavelengths and the corresponding HPhP modes, represented by the IFCs.}
    \label{fig:Ga2O3_HPhP}
\end{figure}

The second example is β-\ce{Ga2O3}, whose permittivity tensor contains nonzero off-diagonal elements due to its monoclinic crystal structure. Such a non-orthogonal permittivity tensor enables shear HPhP~\cite{passler_hyperbolic_2022}, making β-\ce{Ga2O3} a good candidate to demonstrate and analyze $\Lambda_1$, i.e., $45^\circ/135^\circ$ in the $xy$ plane.

The resulting normalized Stokes parameters of β-\ce{Ga2O3} are shown in Fig.~\ref{fig:Ga2O3_Stokes}. Unlike the previous α-\ce{MoO3}, the off-diagonal components of the β-\ce{Ga2O3} permittivity tensor rotate and shear the polariton IFC relative to the coordinate axes. Consequently, the group velocity is no longer aligned with a principal in-plane direction, giving rise to simultaneous nonzero $\Lambda_1$ and $\Lambda_3$, as shown in Fig.~\ref{fig:Ga2O3_HPhP}. $\Lambda_1$ and $\Lambda_3$ reach their maximum amplitudes in the near field around $d = 0.2\text{-}0.5$ μm, depending on the layer thickness. In addition, the amplitude and the decay length of the normalized Stokes components follow the same trend as in the α-\ce{MoO3} case when the slab thickness increases. Compared with the isotropic case, the thermal emissions from anisotropic materials show different polarization features, especially in the near field, due to the support of HPhP modes and anisotropic optical properties.

\begin{figure}[b]
    \centering
    \includegraphics[width=1\linewidth]{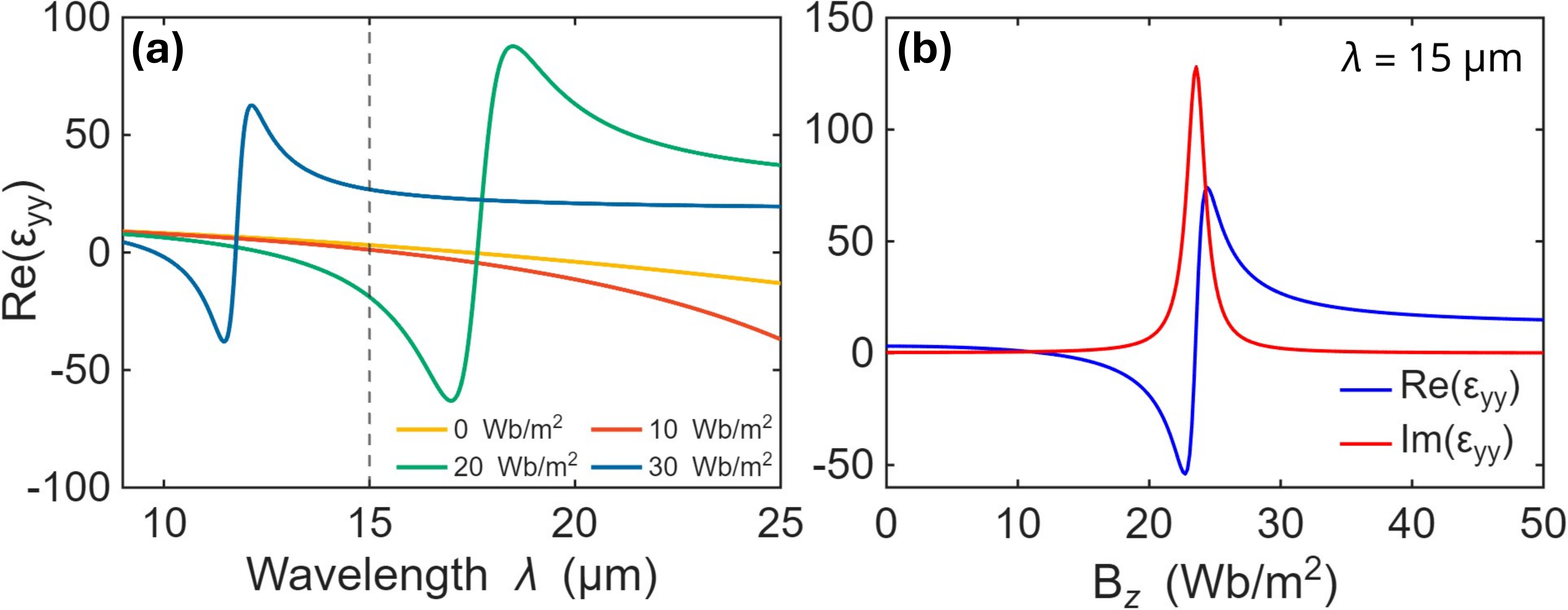}
    \caption{(a) Permittivity spectrum of the n-type doped InAs in the $y$ direction, $\varepsilon_{yy}$, under discrete steps of \textbf{B} fields along $z$ direction, ranging from 0 to 30 $\mathrm{Wb/m^2}$. (b) $\varepsilon_{yy}$ versus magnitude of $B_z$ from 0 to 50 $\mathrm{Wb/m^2}$ at $\lambda = 15$ μm. $\varepsilon_{xx}$ has the same value as $\varepsilon_{yy}$.}
    \label{fig:InAs_eps}
\end{figure}

\subsection{\label{sec:nonreciprocal} Nonreciprocal materials}
For all reciprocal materials considered in the preceding sections, the circular-polarization components vanish, $\Lambda_2=\Lambda_5=\Lambda_7=0$. This behavior is consistent with the absence of equilibrium thermal spin in reciprocal media, which as theoretically demonstrated by Khandekar and Jacob~\cite{khandekar_thermal_2019}. Breaking reciprocity, however, allows nonzero thermal spin and therefore nonvanishing circular polarization in the thermal emission. To demonstrate such a behavior, a nonreciprocal material, n-type doped InAs, is considered as an example, and circular polarization is observed in its thermal emission when the reciprocity is broken under an applied magnetic field. This example shows that circular polarization in the near field can also be captured using the method presented in this work. Furthermore, the results are consistent with the findings of Ref.~\cite{khandekar_thermal_2019} and provide a slightly different perspective.

InAs exhibits strong magneto-optical response when external magnetic field is applied. Consider a planar slab of n-type doped InAs is placed in the same manner as Fig.~\ref{fig:setup}, and uniform magnetic flux density \textbf{B} is applied along the $+z$ direction. The InAs changes its in-plane permittivity from diagonal to antisymmetric with off-diagonal terms due to the external \textbf{B} field. The change of $\varepsilon_{yy}$, for example, is shown in Fig.~\ref{fig:InAs_eps} with the permittivity data from Ref.~\cite{shayegan_nonreciprocal_2022}.

\begin{figure}[t]
    \centering
    \includegraphics[width=1\linewidth]{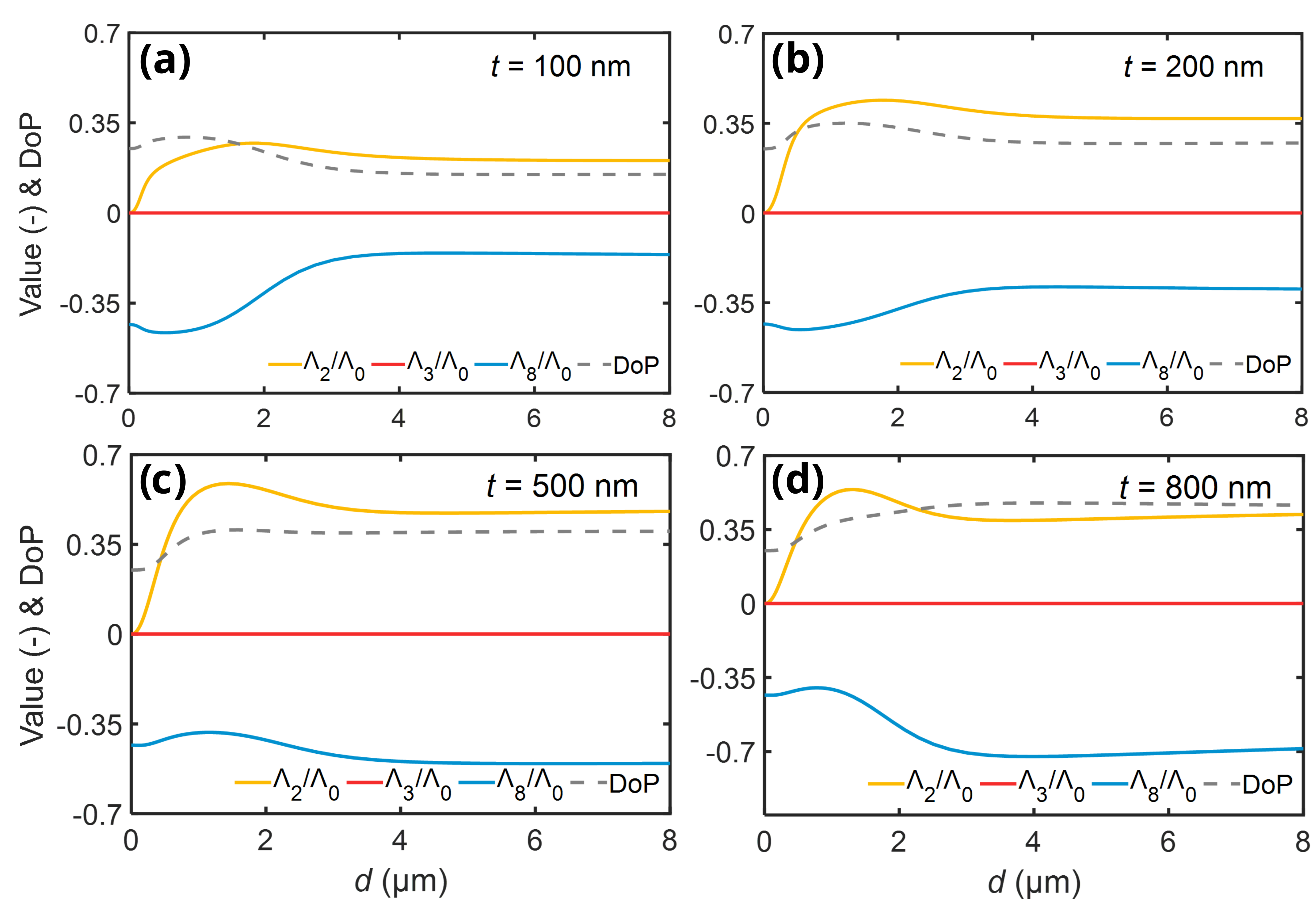}
    \caption{Normalized Stokes parameters of planar doped InAs slabs with different thicknesses $t=$ (a) 100 nm, (b) 200 nm, (c) 500 nm, and (d) 800 nm, at $\lambda=15$ μm. The applied \textbf{B} is 10 $\mathrm{Wb/m^2}$ along the $+z$ direction. The Stokes parameters that are not shown are all zeros.}
    \label{fig:InAS_Stokes}
\end{figure}

The calculated Stokes components exhibit a nonzero positive $\Lambda_2$ over a broad range of distances, indicating a net circular-polarization component in the $xy$ plane, while $\Lambda_1$ and $\Lambda_3$, which are related to in-plane linear polarization, remain zero. The simultaneous vanishing of $\Lambda_1$ and $\Lambda_3$ indicates that the in-plane polarization asymmetry is predominantly associated with the circular-polarization component rather than linear polarization. Similar to other cases in previous sections, the DoP reaches its surface limit value of 0.25 and converges to different far-field limits due to different thicknesses. A distinctive behavior appears as $d\rightarrow0$, where $\Lambda_2$ vanishes. For other Stokes parameters, such as $\Lambda_3$ or $\Lambda_8$, they converge to a finite value when $d$ is close to 0, but $\Lambda_2$ always converges to 0 in the doped InAs case, implying the phase relation between $E_x$ and $E_y$ is broken at such a close distance, which may be explained by the disruption from surface waves with large \textit{k} components.
\begin{figure}
    \centering
    \includegraphics[width=0.65\linewidth]{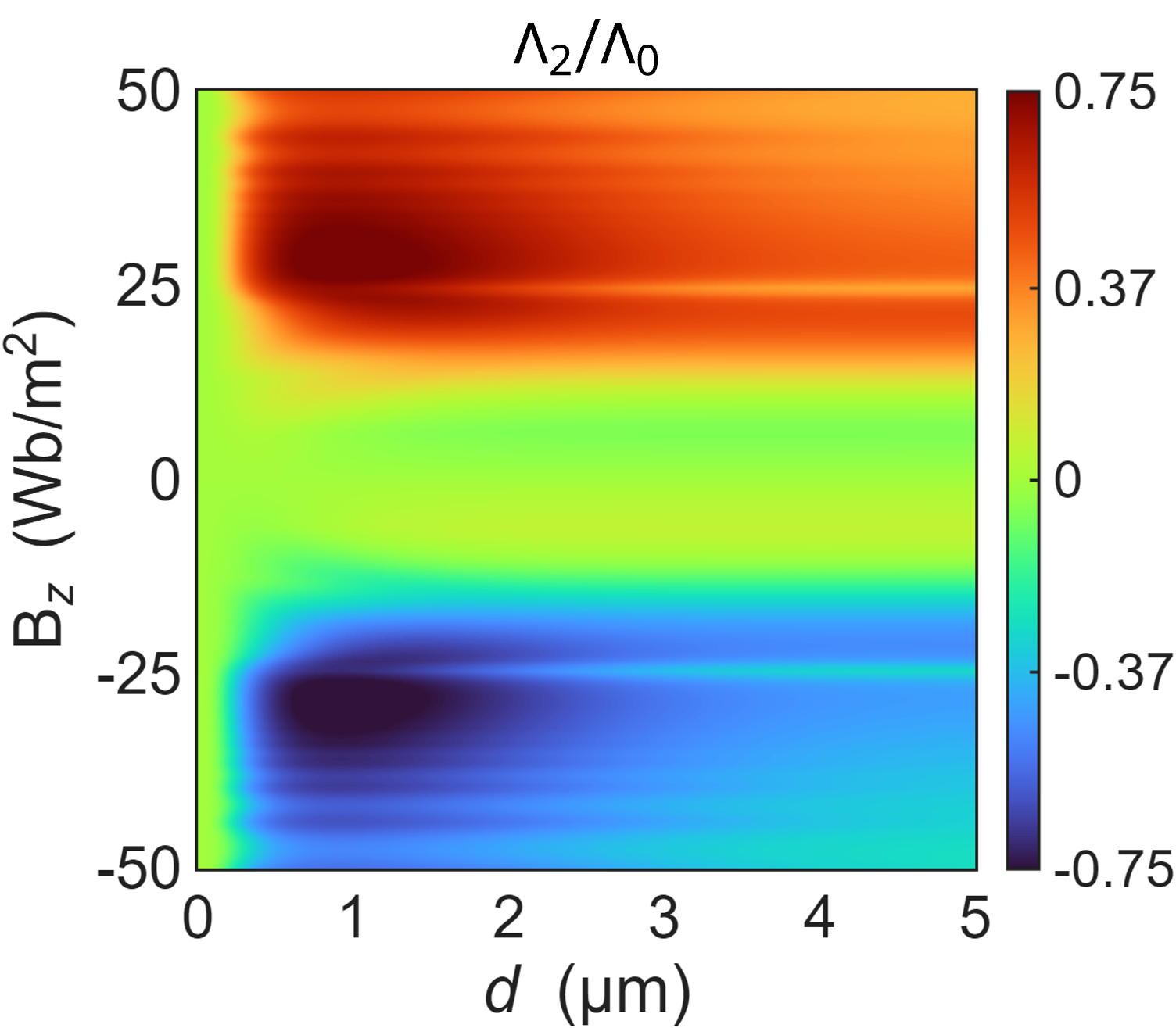}
    \caption{Normalized Stokes parameter $\Lambda_2/\Lambda_0$ under different magnetic bias $B_z$ at distance $d$. The positive and negative values of $B_z$ correspond to the + and $-z$ directions, respectively.}
    \label{fig:InAs_Mirror}
\end{figure}

In addition, the chiral responses of the emitter are investigated under reversed magnetic field bias, where the amplitude of $B_z$ goes from -50 to 50 $\mathrm{Wb/m^2}$, as shown in Fig~\ref{fig:InAs_Mirror}. A negative value of $B_z$ corresponds to the $-z$ direction. The result shows that, for both circular polarizations, the degree of circular polarization reaches its maximum of $\pm0.79$ at $d=0.91$ μm under $B_z=\pm28\ \mathrm{Wb/m^2}$.
The change of sign in $\Lambda_2$ indicates a clear breaking of time-reversal symmetry as the circular polarization changes its handedness under reversed magnetic bias.

\section{\label{sec:conclusion} Conclusion}

In this work, a theoretical method for characterizing the 3D polarization of thermal electromagnetic fields at an arbitrary distance from a thin-film emitter is developed that can be applied to anisotropic materials as well as nonreciprocal materials. Fluctuational electrodynamics is used to determine the $3\times3$ electric-field coherence matrix, from which generalized 3D Stokes parameters and the total degree of polarization are obtained. For isotropic materials, rotational symmetry eliminates the in-plane linear- and circular-polarization components, while an imbalance between the in/out-of-plane field correlations remains. Its near-field behavior is strongly modified by SPhPs, while its far-field behavior is associated primarily with the difference between TE and TM emissivities and with thin-film interference. For anisotropic materials, HPhPs produce additional linear-polarization components. The polarization angles are closely related to the anisotropic polariton dispersion and corresponding directions of group velocity. In the doped InAs example, breaking reciprocity through an external magnetic field gives rise to a nonzero circular-polarization component, demonstrating that the same formalism can describe thermal fields with optical spin. The method developed in this work may be used for analyzing and controlling the polarization in near-field radiative heat transfer, thermal photonics, and other systems in which conventional two-dimensional polarimetry is insufficient.

\begin{acknowledgments}
This work was supported by the Department of Energy (DE-SC0018369) for
X.Z. and Z.M.Z. and the National Science Foundation (DMR-2323909) for
W.C.
\end{acknowledgments}

\bibliography{references}

\end{document}